# MEASURING GALAXY MASSES USING GALAXY-GALAXY GRAVITATIONAL LENSING

Tereasa G. Brainerd[1], Roger D. Blandford[2] & Ian Smail[3]

[1] Los Alamos National Laboratory
Theoretical Astrophysics (T6)
MS B288
Los Alamos, NM 87545

[2] California Institute of Technology
Theoretical Astrophysics 130-33
Pasadena, CA 91125

[3] The Observatories of the Carnegie Institution of Washington
813 Santa Barbara St.
Pasadena, CA 91101

## ABSTRACT

We report a significant detection of weak, tangential distortion of the images of cosmologically distant, faint galaxies due to gravitational lensing by foreground galaxies. A mean image polarisation of $\langle p \rangle = 0.011 \pm 0.006$ (95% confidence bounds) is measured for 3202 pairs of source galaxies with magnitudes in the range $23 < r \leq 24$ and lens galaxies with magnitudes $20 \leq r \leq 23$. The signal remains strong for lens–source separations $\lesssim 90$ arcsec, consistent with quasi-isothermal galaxy halos extending to large radii ($\gtrsim 100 h^{-1}$ kpc). Our observations thus provide the first evidence from weak gravitational lensing of large scale dark halos associated with individual galaxies. The observed polarisation is also consistent with the signal expected on the basis of simulations incorporating measured properties of local galaxies and modest extrapolations of the observed redshift distribution of faint galaxies. From the simulations we derive a best-fit halo circular velocity of $V^* \sim 220$ km s$^{-1}$ and characteristic radial extent of $s^* \gtrsim 100\ h^{-1}$ kpc. Our best-fit halo parameters imply typical masses for the lens galaxies within a radius of 100 $h^{-1}$ kpc on the order of $1.0^{+1.1}_{-0.7} \times 10^{12}\ h^{-1}\ M_\odot$ (90% confidence bounds), in good agreement with recent dynamical estimates of the masses of local spiral galaxies. This is particularly encouraging as the lensing and dynamical mass estimators rely on different sets of assumptions. Contamination of the gravitational lensing signal by a population of tidally distorted satellite galaxies can be ruled out with reasonable confidence. The prospects for corroborating and improving this measurement seem good, especially using deep HST archival data.

# 1. INTRODUCTION

The notion that cosmologically-distributed masses might cause weak but measurable changes to the shapes of distant galaxies has a long history. The most striking example of this "cosmological distortion effect" (Kristian & Sachs 1966) is the distortion of distant galaxies into giant arcs caused by rich clusters of galaxies (eg. Fort & Mellier 1994). Attempts have also been made to measure the weak distortion of distant galaxy images by mass fluctuations associated with large-scale structure (eg. Kristian 1967; Valdes, Tyson & Jarvis 1983; Mould et al. 1994, Paper I). These studies have progressed to the point where they can produce interesting limits on the large scale distribution of mass in the universe. The purpose of this paper, however, is to show how weak gravitational lensing can be used to study the mass distributions on much smaller scales, those associated with individual galaxies.

Galaxies constitute the fundamental building blocks of luminous structure in the universe, yet we are largely ignorant of such basic physical parameters as their typical masses and radial extents. Popular theories of galaxy formation predict that most bright galaxies should reside in massive ($\gtrsim 10^{12} M_\odot$) dark halos that extend far beyond their optical radii ($\gtrsim 100 h^{-1}$ kpc). Direct observational evidence to test such theories is, however, scarce. While observations of the central parts of galaxies provide good mass estimates for these regions, the lack of information on the form and extent of the dark halos of individual galaxies limits our determination their masses. Observations of local galaxies, most importantly our own (eg. Fich & Tremaine 1991), indicate that the majority of bright spiral galaxies have dark halos which extend isothermally out to at least $\sim 30$ kpc. Studies using the dynamical properties of ensembles of faint companions to samples of bright spirals favour a continuation of the halo out to radii $\sim 100$ kpc (Zaritsky & White 1994). On purely theoretical grounds, if we suppose the mass of the halos increases linearly with radius out until their density reaches the critical density, they will have outer radii $\sim (2/3)^{1/2}(V_c/H_0) \sim 1 - 2h^{-1}$ Mpc, where $V_c$ is the circular velocity. Alternatively, the halos of most giant galaxies may be truncated at radii smaller than a tenth of this value.

In this paper we will report on a significant detection of distortion of the shapes of distant galaxy images due to weak gravitational lensing by individual foreground galaxies. Using the observed gravitational lensing signal we will then attempt to measure masses for typical field galaxies. The advantages of the lensing approach to determining galaxy masses are two-fold. Firstly, it is capable of probing the halos of galaxies out to very large radii, $r \gtrsim 100 h^{-1}$ kpc, where few classical techniques are viable. Secondly, the lensing analysis is relatively unaffected by the dynamical properties of the possibly unvirialised outer regions of the halos. Lack of detailed information about the dynamics at large radii in galaxies is one of the central problems for the application of dynamical mass estimators. The dynamical and lensing mass estimates depend upon different model assumptions and a comparison of the results obtained from the two techniques is one of the few ways in which the validity of these assumptions can be tested.

The signal for which we are searching is a distortion of the images of faint galaxies resulting in a weakly preferred alignment of faint galaxies around brighter galaxies. If the faint galaxies are gravitationally lensed by the brighter systems, the major axes of their images will tend to lie perpendicular to the radius vectors joining the faint galaxies to the lens centres (Fig. 1). The strength of this signal depends upon the distances of the lens and source galaxies, the mass of the lens, and the angular separation of the lens and source on the sky. With statistical information from direct spectroscopic studies (eg. Lilly 1993, Tresse et al. 1993) on the redshift distributions of both the source and lens galaxy samples we can therefore solve for the typical masses of the lensing galaxies.

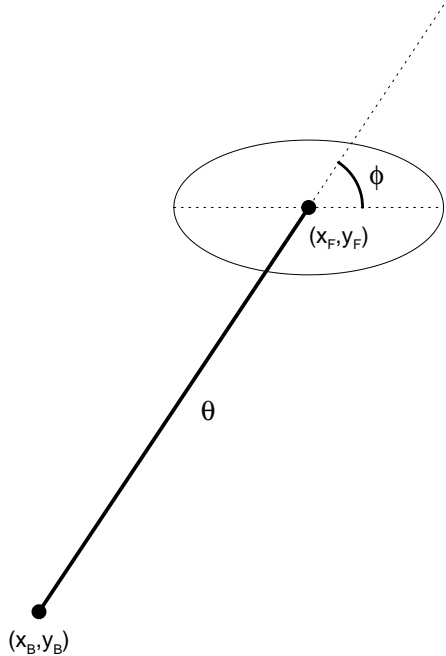

Fig. 1: Orientation of faint galaxies relative to bright galaxies

The detection of galaxy–galaxy lensing was first attempted by Tyson et al. (1984) using scans of photographic plates (cf. also Webster 1983) from which they tried to measure an excess of background galaxy images tangentially elongated with respect to brighter, candidate lens galaxies. The galaxy sample available to Tyson et al. was very large, $\sim 47,000$ background galaxies with $22.5 < J < 23.5$ and $\sim 12,000$ lens galaxies with $19 < J < 21.5$. Faint image position angles were measured for over $28,000$ galaxy pairs and no statistically significant difference from an isotropic distribution was seen for galaxy separations greater than $\sim 3$ arcsec. On this basis, Tyson et al. concluded that the typical galaxy circular velocity had a surprisingly small upper bound, $\lesssim 170$ km s$^{-1}$. Unfortunately, the relatively modest seeing of their plate material could have a very detrimental effect on the efficiency of the test. Moreover, Kovner & Milgrom (1987) performed a more careful calculation of the magnitude of the effect anticipated, taking into account integration over galaxy luminosity functions and distances as well as the correlation of internal, galaxian velocity dispersions with luminosity and concluded that the observation of Tyson et al. was consistent with conventional dynamical models of local galaxies.

In §2 we describe our observations and present our results for the detection of weak gravitational lensing of distant galaxy images by foreground galaxies, together with the analyses we have performed to attempt to understand the importance of systematic errors. In §3 we describe model calculations, both analytic and Monte Carlo, that we have used to translate our measured galaxy–galaxy lensing signal into a quantitative statement about galaxy masses and extents, results of which are presented in §4. Finally, in §5 we relate our measurement of galaxy masses and radial extents to more conventional determinations of these quantities.

## 2. MEASUREMENT OF GALAXY-GALAXY LENSING

To estimate roughly the expected strength of the galaxy–galaxy gravitational lensing signal, let us model a lens galaxy as a singular isothermal sphere with circular velocity $V_c$. An ellipticity $\sim 2\pi V_c^2/c^2\theta$ will then be induced in the image of a background faint galaxy located an angular distance $\sim \theta$ from the lens. This is of the order of a few percent effect for galaxy pairs with separations $\theta \sim 30$ arcsec where the lens galaxy is a typical bright spiral galaxy. In order to detect this signal in the presence of the noise associated with the intrinsic galaxy shapes, over a thousand foreground–background galaxy pairs must be measured. If a sufficiently large number of pairs are available, it may also be possible to use the dependency of the lensing signal on the distance from the lens centre, $\theta$, to study the angular extent of galaxy halos. In the following section we discuss an observational dataset which should be of sufficiently quality, depth and size to allow us to detect galaxy–galaxy lensing.

### 2.1 Observational Data

The imaging data used in our analysis is of a single $9.6 \times 9.6$ arcminute blank field centred on $\alpha(1950) = 17^h21^m07^s$ $\delta(1950) = +49°52'21''$, taken in Gunn $r$. The data were acquired in periods of good seeing, 0.7–0.9 arcsec, using the direct imaging mode of the COSMIC imaging spectrograph (Dressler et al. 1995) on the 5-m Hale telescope, Palomar. These data have been used previously to study the coherent distortion of faint galaxy images due to weak gravitational lensing by large-scale structure (Paper I) and the angular clustering statistics of faint galaxies (Brainerd, Smail, & Mould 1995). The reduction of the data to a catalogue of detected objects is detailed in Paper I.

The final stacked $r$ image used for our principal data analysis consists of a total of 24.0 ksec integration, has a $1\sigma$ surface brightness limit of $\mu_r = 28.8$ mag arcsec$^{-2}$, seeing of 0.87 arcsec FWHM, and a total area of 90.1 arcmin$^{-2}$. The object catalogue created from this frame using the FOCAS image analysis package (Valdes 1982) contains $\sim 6600$ objects brighter than the 80% completeness limit of $r = 26.2$. Adopting a conservative magnitude limit of $r \leq 26.0$, where the detections are approximately 97% complete, we obtain a cumulative surface density of 71.8 galaxies arcmin$^{-2}$ or $2.6 \times 10^5$ deg$^{-2}$. Due to the presence of classical distortion in the corners of the frame, our analysis uses only those galaxies which lie within a circle of radius $4.8'$, centred on the chip. There are 4819 galaxies within this area brighter than $r = 26.0$.

An additional, shallower $g$-band image of the same field is also available and is used solely for the purpose of providing colour information on the objects detected in the deep $r$ image. The $g$-band image has a total exposure time of 6.0 ksec and seeing of 1.2 arcsec.

The median colour errors for galaxies detected on the $r$ image with $(g-r) \leq 1$ at $r \sim 26$ (approximately 80% of the total population) are $\Delta(g-r) \sim 0.2$.

In order to calculate the gravitational lensing signal yielded by our models and to estimate the lensing-induced mean image polarisation of our faint objects, we shall require the distribution function of the intrinsic source galaxy ellipticities. To linear order the source ellipticity distribution can be estimated by the ellipticity distribution of the images. We have, therefore, measured the distribution function of image ellipticities in our sample and find that it is adequately fit by the normalized distribution function

$$P_\epsilon(\epsilon) = 64\epsilon \exp[-8\epsilon] \qquad (2.1)$$

with mean ellipticity $\langle \epsilon \rangle = 0.12$ and we estimate the error to be $\pm 0.02$.

## 2.2 Position angle probability distribution

In this section we investigate the orientation of faint galaxies relative to the directions to nearby bright galaxies (see Fig. 1). The unweighted second moments of the intensity were measured and a complex orientation, $\chi$, formed for each faint image. The modulus of $\chi$ is $(a^2 - b^2)/(a^2 + b^2)$ where $a^2$ and $b^2$ are the principal second moments of the intensity provided by FOCAS. (See Paper 1 for a discussion of alternative prescriptions for describing the image shapes.) For $a \sim b$, $|\chi|$ is approximately equal to the measured ellipticity, $\epsilon = 1 - b/a$, and this identification is adequate for our purposes. The phase of $\chi$ is twice the position angle, $\phi$, and we shall use the convention of measuring $\phi$ to be the angle between the major axis of the equivalent ellipse and the radius vector, measured in a counter-clockwise sense. We also combine positive and negative position angles so that $\phi$ is restricted to $[0, \pi/2]$.

In the absence of distortion of the faint galaxy images, we expect a uniform distribution of their position angles for all projected separations, $\theta$, between the candidate lenses and sources. When the background galaxy images are gravitationally lensed by foreground galaxies, we expect a distribution of position angles that is non-uniform with a deficit of faint images oriented radially ($\phi = 0$) and an excess of faint images oriented tangentially ($\phi = \pi/2$). As we show in §3, the deviation of the distribution from uniform should exhibit a $\cos 2\phi$ variation. At large projected separations, of course, the distribution of $\phi$ should become uniform.

In Fig. 2 we show the observed faint galaxy orientation distribution, $P_\phi(\phi)$, for lens–source projected separations of $5 \leq \theta \leq 34$ arcsec. To determine $P_\phi(\phi)$ we use the 439 bright galaxies with magnitudes in the range $20.0 \leq r_d \leq 23.0$ (the candidate lenses) and faint galaxies with magnitudes in the ranges: (a) $23.0 < r_s \leq 24.0$, (b) $23.0 < r_s \leq 25.0$ and (c) $23.0 < r_s \leq 26.0$. The magnitude limits used to select the candidate lens galaxies are roughly equivalent to the depth of current faint spectroscopic surveys (eg. Lilly 1993, Tresse et al. 1993), which provide statistical distance information on this population. The number of faint images and faint–bright pairs as a function of limiting source magnitude are summarised in Table 1. The lower limit on $\theta$ was chosen so as to avoid overlapping faint and bright image isophotes. We discuss the choice of upper limit on $\theta$ below; the value here corresponds to an average impact parameter of $\sim 120 h^{-1}$ kpc at our median lens redshift. The error bars on $P_\phi(\phi)$ are estimated from bootstrap resampling of the data.

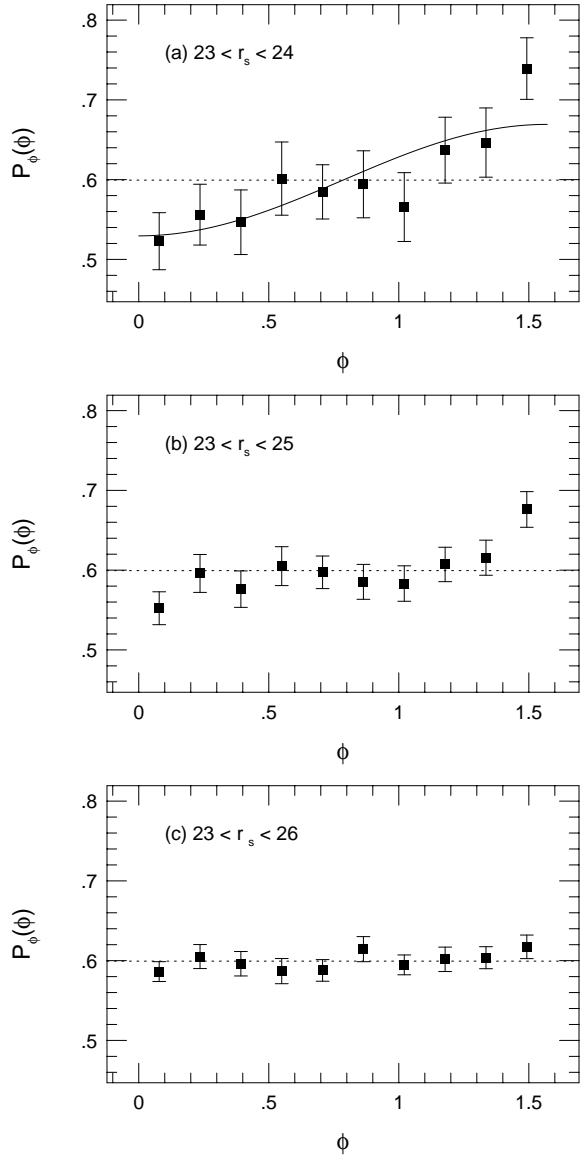

Fig. 2: Probability distribution $P_\phi(\phi)$ of orientation of faint galaxies relative to the directions of bright galaxies with projected separations $5 \leq \theta \leq 34$ arcsec. The bright galaxies have $20 \leq r \leq 23$ and the faint galaxies have magnitudes in the ranges indicated in the text and figure panels. For the best case, (a), of a non-uniform $P_\phi(\phi)$, the best fitting theoretical $\cos 2\phi$ variation is also shown.

From Fig. 2, the (binned) distribution of faint image position angles appears non-uniform for the two brighter galaxy samples. In addition, $P_\phi(\phi)$ for the two brighter samples is qualitatively in agreement with what one would expect in the case of lensing by the bright foreground galaxies. We investigate the significance of the deviation of $P_\phi(\phi)$ from a uniform distribution using standard statistical techniques.

A $\chi^2$ test performed on the binned $P_\phi(\phi)$ in Fig. 2 rejects a uniform distribution

($P_\phi(\phi) = 2/\pi$) at a modest confidence level ($> 97\%$) for both of the brighter samples and cannot reject a uniform distribution for the faintest sample. The sensitivity of the $\chi^2$ test to the arbitrary choice of bin size when using a continuous variable is well known. With this in mind, we consider the Kolmogorov–Smirnov (KS) test, which compares the observed continuous, *cumulative* distribution with a similar theoretical distribution, to be a better statistic for our purposes. KS tests performed on the cumulative $P_\phi(\phi)$ distributions corresponding to the binned distributions shown in Fig. 2 reject uniform distributions at better than the 99% confidence level for both brighter source samples, however we still can not rule out the hypothesis that our faintest sample is uniformly distributed. A summary of rejection confidence levels as a function of the source limiting magnitude is given in Table 1.

We do not view the absence of an obvious lensing signal in our faintest source sample as a strong concern. In a universe where galaxies at fainter magnitudes are observed at progressively more distant epochs we might expect the lensing signal to continue to strengthen in fainter samples. Unfortunately, as we reach fainter in our Universe the typical angular size of galaxies decreases, and by $r \sim 26$ they have half-light radii of $\lesssim 0.4$ arcsec. This is small compared to our seeing and means that we only partially resolve our faintest galaxies, leading to increasing errors in the determination of their ellipticities and position angles, degrading the gravitational lensing signal that we measure. Such a trend is clear from the data exhibited in Fig. 2. We must, therefore, settle for an optimal choice of limiting magnitude for the faint images and, from Fig. 2, this appears to be $r \sim 24$ for our data. We therefore adopt a source sample defined by $23.0 < r_s \leq 24.0$ (Fig. 2a) for our principal analysis.

As we anticipate $P_\phi(\phi)$ will exhibit a $\cos 2\phi$ variation in the case that the faint images have indeed been gravitationally lensed by the brighter galaxies, we show in Fig. 2a the best fitting $\cos 2\phi$ variation for $P_\phi(\phi)$ for the subsample that will be used for our principal analysis.

### 2.3 Possible systematic effects

We have performed a number of tests to investigate possible systematic effects in our data which would give rise to the observed non-uniform $P_\phi(\phi)$. To begin with we considered lens–source pairs with projected separations of $5 \leq \theta \leq 34$ arcsec, lens galaxies with $20 \leq r_d \leq 23$, and source galaxies with $23 < r_s \leq 24$, and computed $P_\phi(\phi)$ for the following:

(1) $\phi$ taken to be the position angle of the *lens* galaxies relative to the lines connecting their centroids with those of the sources
(2) the position angle of the source images relative to lines connecting their centroids with random points (i.e. not corresponding to the centroids of the lenses)
(3) the position angle of the source images relative to the lines connecting their centroids to bright stars on the frame
(4) random position angles were substituted for the observed position angles of the faint images
(5) using the observed centroids and source position angles, the distributions of positive and negative values of $\phi$ were computed independently and compared

In cases (1)-(4), neither the $\chi^2$ nor the KS test rejects the uniform distribution (i.e. no "signal" is observed in these cases). In case (5), the distributions obtained for the two subsamples are statistically indistinguishable and the KS test rejects uniform distributions at the $\sim 99\%$ confidence level (a somewhat weaker rejection than that obtained when positive and negative values of $\phi$ were combined, cf. Table 1).

Next we discuss the possible role of a point spread function (psf) asymmetry in creating the observed signal. The presence of a non-circular psf or guiding errors introduces a preferred orientation in the object images and can give rise to a non-uniform $P_\phi(\phi)$ on very large angular scales. When the orientation of the faint galaxies relative to the bright galaxies is computed on small scales ($\lesssim 1/4$ the size of the frame), this effect is canceled out due to the fact that we can compute the orientations of the faint images in complete annuli around the majority of the bright centers. However, on scales $\gtrsim 1/2$ the size of the frame, the effect becomes very significant as we can no longer compute the orientations of the faint images in complete annuli around most of the centers. In general, the edge effect at large scales that gives rise to the non-uniform $P_\phi(\phi)$ will not resemble that expected for a gravitational lensing signal, being of a larger amplitude and peaking at an angle $\phi$ corresponding to a combination of the preferred orientation of the faint images and their direction vectors relative to the bright centers.

In the case of our data, there is a slightly elliptical psf measured from the bright stars on the frame and a corresponding weakly preferred image orientation. Computing $P_\phi(\phi)$ for our fiducial subsamples of galaxies, we find it is consistent with a uniform distribution on scales of $\sim 100$–$150$ arcsec (as would be expected for a gravitational lensing signal), but on scales $\gtrsim 250$ arcsec, $P_\phi(\phi)$ is significantly non-uniform due to the edge effect. The form of $P_\phi(\phi)$ on these scales, unlike Fig. 2, does not coincide with the expectations of a gravitational lensing signal at these scales, being of a larger amplitude and peaking at $\phi = \pi/4$ with corresponding suppressions at $\phi = 0$ and $\phi = \pi/2$. Such a signal (both amplitude and shape) in our data is expected at this scale given the observed preferred orientations of the objects (mean position angle of $\sim -5\pi/18$).

Finally, as a check for systematics associated with the image detection routines and second moment determinations, $P_\phi(\phi)$ was computed from a simulated data frame in which a signature of weak gravitational lensing was not included. Images on the simulated frame were assigned random locations and orientations. The galaxy parameters (ellipticities, scale sizes, and magnitudes) were chosen such that after the images were convolved with the telescope psf, the distribution of these parameters matched the observed distributions. The surface density of objects above the completeness limit was matched to that observed and, in addition, fainter galaxies were added by extrapolating the number counts 2 magnitudes fainter in order to simulate the effects of crowding and merging of these undetected faint galaxies on the detected objects. The pixellated images were convolved with the psf defined by the telescope and atmosphere, and sky noise was added to obtain the same detection limits as in the observed images. The simulated frame was then analyzed using FOCAS and a catalogue of objects produced using the same procedure implemented for the actual dataset. Using the parameters corresponding to Fig. 2a ($5 \leq \theta \leq 34$ arcsec, $20 \leq r_d \leq 23$, $23 < r_s \leq 24$), $P_\phi(\phi)$ for the simulated faint galaxy images was computed. Again, neither the KS nor the $\chi^2$ test rejects the uniform distribution for the artificial data frame and we conclude that our observed non-uniform $P_\phi(\phi)$ is not a result of errors in the analysis.

From the results of our various tests we therefore conclude that the signal observed in Fig. 2 is real and not an artifact of the dataset. In the following section we develop a model to recreate the observed non-uniform $P_\phi(\phi)$ using galaxy–galaxy gravitational lensing.

## 3. SIMULATIONS OF GALAXY-INDUCED IMAGE DISTORTIONS

In order to compare the magnitude of our apparent gravitational lensing signal with theoretical expectation, we have carried out a variety of analytic and Monte Carlo simulations of weak lensing of galaxy images. To do this we adopt a simple, fiducial galaxy model, compute the expected signal using analytic and Monte Carlo simulations, then explore the sensitivity of the predictions to changes in our assumptions. In so doing, we are implicitly assuming that the majority of our lens galaxies form a one parameter family as far as their gravitational properties outside $\sim 30$ kpc are concerned. Some cosmogonic theories posit that the outer parts of ellipticals are similar to the halos of spirals (eg. de Zeeuw & Franx 1991), a view that is supported by X-ray observations (eg. Fabbiano 1989). We shall therefore not distinguish the minority of ellipticals from the spirals present in our lens sample.

### 3.1 Galaxy mass model

In devising an appropriate galaxy model, two simplifications present themselves. Firstly, we are quite unconcerned with the details of the galaxy cores as we expressly exclude $\theta \lesssim 5$ arcsec. We can therefore adopt a model galaxy potential that is singular as $r \to 0$. Secondly, almost all of our images are weakly distorted as they lie well outside the tangential critical curves formed by the galaxies. This implies that if the ellipticity in the potential is $\epsilon_g(r)$ at radius $r$, the average image polarisation $\langle p \rangle (r, \epsilon_g)$ will differ from that of a circular lens with the same mass contained within a circle of radius $r$ by an amount $O(\epsilon_g^2 \langle p \rangle (r, 0))$. As $\epsilon \lesssim 0.3$ for the bulk of the images, this correction is small compared with the uncertainty implicit in the model. We therefore treat the galaxy potential as circularly symmetric. It should be emphasized at this stage that it is not at all likely that the outer parts of galaxies will have had time to relax completely into the ellipsoidal shapes associated with the luminous inner regions. In particular, the largest equipotentials associated with individual galaxies may be quite non-circular. However, as weak lensing depends linearly upon the mass distribution and we are only attempting a statistical measurement, our model corresponds to a circularly-averaged mass distribution. Note that when an image is distorted by two or more distinguishable lens galaxies, their linear effects can be superposed.

After some experimentation, we have found that a simple model mass distribution for the dark matter halos which contains an outer characteristic scale, $s$, as a free parameter is

$$\rho(r) = \frac{V_c^2 s^2}{4\pi \mathcal{G} r^2 (r^2 + s^2)} \tag{3.1}$$

where $V_c$ is the de-projected circular velocity for $r \ll s$ and $\mathcal{G}$ is Newton's constant. For $r \gg s$, the density declines as $r^{-4}$ and the polarisation signal becomes small.

The surface density associated with this space density is given by

$$\Sigma(R) = \int_{-\infty}^{\infty} dx \rho[(R^2 + x^2)^{1/2}] = \frac{V_c^2}{4\mathcal{G}R}[1 - (1 + X^{-2})^{-1/2}] \tag{3.2}$$

where $X \equiv R/s$. The total mass is finite

$$M = \frac{\pi s V_c^2}{2\mathcal{G}} \tag{3.3}$$

and the mass contained within a radius, $r$, is given by

$$M(r) = \frac{V_c^2 s}{\mathcal{G}} \mathrm{Tan}^{-1}(r/s). \tag{3.4}$$

We shall need the 2-dimensional potential, relative to the potential at some large but finite radius, $R_L$

$$\Phi(R) = -2 \int_R^{R_L} dR' \frac{M(R')}{R'} = \frac{-\pi V_c^2 s}{\mathcal{G}} \left[(1+X^2)^{1/2} - X - \ln[(1+X^2)^{1/2} + 1]\right] \tag{3.5}$$

where we drop a constant, logarithmic contribution from the outer reference radius which does not contribute to the deflection and the polarisation. In order to demonstrate that this potential is physically realisable, we compute an associated distribution function in Appendix A and demonstrate that it is stable to small perturbations.

### 3.2 Image polarisation

Using the model potential for an individual lens galaxy given by Eq. (3.1), we now compute the deflection angle of light rays and the resultant image polarisation. Let the lens galaxy have angular diameter distance $D_d$ and source galaxy have angular diameter distances $D_s, D_{ds}$ as seen from Earth and the lens, respectively. The deflection of a ray from the source is then given by

$$\alpha(X) = \frac{2D_{ds}}{D_s c^2} \frac{d\Phi}{dR} = \frac{2\pi V_c^2 D_{ds}}{D_s X c^2}[1 + X - (1 + X^2)^{1/2}] \tag{3.6}$$

(eg. Blandford & Narayan 1992; Schneider, Ehlers & Falco 1993). Equating the image polarisation, $p(X)$, to the induced ellipticity we have

$$p(X) = -\frac{D_d X}{s} \frac{d}{dX}\left(\frac{\alpha(X)}{X}\right) = p_0 G(X), \tag{3.7}$$

where a positive polarisation corresponds to a net tangential elongation of the images. The coefficient $p_0$ is given by

$$p_0 = \frac{2\pi V_c^2 D_d D_{ds}}{s D_s c^2} = \frac{4\mathcal{G} M D_d D_{ds}}{s^2 D_s c^2} \tag{3.8}$$

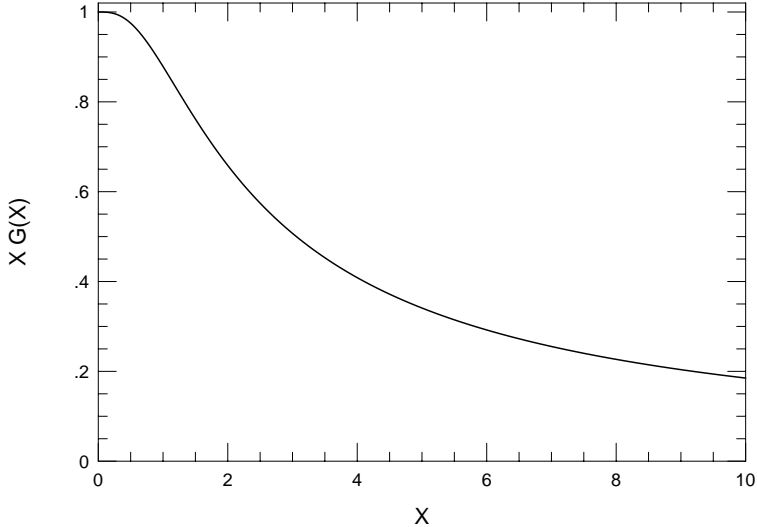

Fig. 3: Scaled polarisation variation with angle $\theta$ for a given galaxy. $G(X)$ is defined in Eq. (3.9) and $X = R/s$.

and the function $G(X)$ is

$$G(X) = \frac{(2+X)(1+X^2)^{1/2} - (2+X^2)}{X^2(1+X^2)^{1/2}}. \qquad (3.9)$$

For $X \ll 1$, $G(X) \sim 1/X$, whereas for $X \gg 1$, $G(X) \sim 2/X^2$. Note also that $G(1) = 0.88$ so that the polarisation $p \sim p_0$ when $R \sim s$. $G(X)$ is plotted in Fig. 3. Note that Eq. (3.6), (3.7) allow us, in principle, to invert a measurement of $p(\theta)$ to obtain the average galaxy potential. However our data are too sparse and our knowledge of the redshift distributions and the luminosity variation of galaxy properties is too incomplete to make this practical.

In order to use our model above to determine the expected image polarisation for galaxies in an observational sample, we need to relate the luminosities of galaxies to the depths of their potential wells. To do this, we use the observations that the local spiral galaxy population can be well-described by a luminosity function scaled to a characteristic luminosity $L^*$ (eg. Loveday et al. 1992) and that these spirals appear to obey a Tully-Fisher relation that relates their luminosities, $L$, to their circular velocities, $V_c$ (eg. Aaronson & Mould 1983). With this in mind, we now introduce two scaling laws roughly consistent with these observations.

Firstly, we assume that the circular velocity, $V_c$, scales as the fourth root of the total luminosity in a given band, in agreement with the Tully-Fisher relation (and also with the Faber-Jackson and fundamental plane laws for ellipticals if we treat $V_c/\sqrt{2}$ as the central velocity dispersion [cf. de Zeeuw & Franx 1991]). Introducing a scaling circular velocity,

$V^*$, we have
$$\frac{V_c}{V^*} = \left(\frac{L_\nu}{L_\nu^*}\right)_r^{1/4} \qquad (3.10)$$
where the subscript $r$ refers to the $r$-band and we assume that the emitted spectrum does not change with cosmological epoch. A fiducial estimate of $V^*$ is 220 km s$^{-1}$ (eg. Fich & Tremaine 1991).

Our second scaling relation is more of a hypothesis. We assume that the total mass to light ratio of a galaxy is a constant independent of its luminosity. In other words we suppose that there is a one parameter family of galaxy potentials whose outer radii, $s$, scale as $s \propto M^{1/2} \propto L_\nu^{1/2} \propto V_c^2$. Introducing a scaling radius, $s^*$, we have
$$\frac{s}{s^*} = \left(\frac{L_\nu}{L_\nu^*}\right)_r^{1/2} = \left(\frac{M}{M^*}\right)^{1/2}. \qquad (3.11)$$

Imposing our galaxy scaling laws above, then, we conclude that $p_0$ does not depend explicitly upon the galaxy mass and we can therefore write
$$p_0 = p^* \frac{D_{ds} D_d H_0}{cD_s} \qquad (3.12)$$
where
$$p^* = 0.10 \left(\frac{V^*}{220\text{km s}^{-1}}\right)^2 \left(\frac{s^*h}{100\text{kpc}}\right)^{-1} \qquad (3.13)$$
is a reference polarisation. The total mass of an $L^*$ galaxy in this model is
$$M^* = 1.8 \times 10^{12} \left(\frac{V^*}{220\text{km s}^{-1}}\right)^2 \left(\frac{s^*}{100h^{-1}\text{kpc}}\right) M_\odot. \qquad (3.14)$$

We now compute the expected mean image polarisation from an observational investigation such as the one that we have described in §2. For simplicity, in our fiducial model we adopt an Einstein-De Sitter universe with angular diameter distances
$$D_d = \frac{2c}{H_0} a_d (1 - a_d^{1/2}), \quad D_s = \frac{2c}{H_0} a_s (1 - a_s^{1/2}), \quad D_{ds} = \frac{2c}{H_0} a_s (a_d^{1/2} - a_s^{1/2}) \qquad (3.15)$$
and expansion factor $a_{d,s} = (1 + z_{d,s})^{-1}$.

We must also allow for a spectral (or "K") correction, which can be fairly large in the $r$-band since the spectrum is quite steep at wavelengths blueward of this band (i.e. $\sim 6500$ Å) in intermediate redshift spirals. As we believe our lenses have redshifts $\sim 0.2 - 0.8$ (see below), we are concerned with spectral shapes in the range $\sim 3500 - 5500$ Å. Comparing with colour measurements of faint galaxies, we find that $\alpha \equiv -d\ln L_\nu / d\ln \nu \sim 3$, very approximately. For ellipticals, the effective value of the spectral index, $\alpha$, is closer to 5 and so they should be somewhat rarer in an $r$-selected faint galaxy sample than they are locally. We also ignore the large dispersion in the colours of observed galaxies and treat them as a one parameter family. We can then relate the apparent magnitude to the luminosity adopting $-18.5 + 5\log h$ as the absolute $r$ magnitude of an $L^*$ galaxy. With this we obtain

$$\frac{L_\nu}{L_\nu^*} = \left(\frac{H_0 D_d}{c}\right)^2 (1+z)^{3+\alpha} 10^{0.4(23.9-r)} \qquad (3.16)$$

at a given emitted frequency.

In order to proceed further, we need redshift distribution functions for the source and lens galaxies as a function of apparent magnitude. Lilly (1993) and Tresse et al. (1993) have presented I-selected redshift surveys which show that galaxies in a magnitude range comparable to that under consideration here ($I \sim 22$ or $r \sim 23$) have a broad redshift distribution extending out to $z \sim 1$. As the results of our simulation are quite sensitive to the form of the redshift distribution used, we adopt a parameterised and normalised redshift distribution

$$F(z,r) = \frac{\beta z^2 e^{-(z/z_0)^\beta}}{\Gamma(3/\beta) z_0^3}. \qquad (3.17)$$

For our fiducial model, we adopt $\beta = 1.5$. The mode is $1.2 z_0$, the median is $z_m = 1.4 z_0$ and the mean is $1.5 z_0$. More generally, we write

$$z_0 = k_z [z_m + z_m'(r-22)]; \quad 20 < r < 24 \qquad (3.18)$$

where the constant $k_z = 0.7$ for $\beta = 1.5$, and the derivative of the median redshift with respect to $r$ magnitude is $z_m' = 0.1$ fiducially.

Next, source galaxies with apparent magnitude $r_s$ are selected within an annular ring of outer radius $\theta_{\max}$, and inner radius $f\theta_{\max}$. We now average the function $G(X)$ over this ring for lenses of fixed redshift $z_d$ and magnitudes $r_d$, obtaining

$$\begin{aligned}\bar{G}(z_d, r_d) &= \frac{\int_{D_d f \theta_{\max}/s}^{D_d \theta_{\max}/s} dX\, X\, G(X)}{\int_{D_d f\theta_{\max}/s}^{D_d \theta_{\max}/s} dX\, X} \\ &= \frac{2S}{(1-f^2)} \left\{ (1-f) - (1+S^2)^{1/2} + (f^2+S^2)^{1/2} \right. \\ &\quad \left. + 2S \ln\left[\frac{S + (1+S^2)^{1/2}}{S + (f^2+S^2)^{1/2}}\right] \right\}\end{aligned} \qquad (3.19)$$

where $S = s/D_d \theta_{\max}$. For $S \ll 1$, $\bar{G} \propto S^2$; for $S \gg 1$, the relation becomes asymptotically, $\bar{G} \sim 1.7 S$.

We next integrate over the source redshift for a fixed lens.

$$\langle p \rangle (z_d, r_d, r_s) = \left(\frac{p^* H_0}{c}\right) \bar{G}(z_d, r_d) D_d \int_{z_d}^\infty dz_s F(z_s, r_s) \left(\frac{D_{ds}}{D_s}\right). \qquad (3.20)$$

Finally, we integrate over the redshift distribution of lens galaxies of magnitude $r_d$ to obtain

$$\langle p \rangle (r_d, r_s) = \left(\frac{p^* H_0}{c}\right) \int_0^\infty dz_d \bar{G}(z_d, r_d) D_d(z_d) F(z_d, r_d) \int_{z_d}^\infty dz_s F(z_s, r_s) \left(\frac{D_{ds}}{D_s}\right). \qquad (3.21)$$

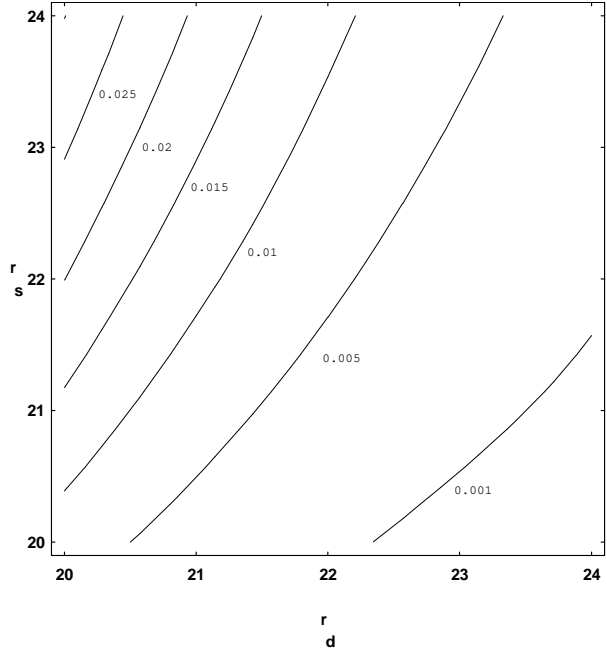

Fig. 4: Theoretical variation of polarisation as a function of source magnitude $r_s$ and lens galaxy magnitude $r_d$ according to Eq. (3.21).

Note that the polarisation is directly proportional to the assumed velocity dispersion of an $L^*$ galaxy but depends non-linearly on a single galaxy structural parameter, $s^*$.

### 3.3 Fiducial lens model

To summarise, our fiducial lens model in an Einstein-De Sitter universe uses:

$$\begin{aligned}
V^* &= 220 \text{kms}^{-1} \\
s^* &= 100 h^{-1} \text{kpc} \\
\alpha &= 3 \\
\theta_{\max} &= 34 \text{ arcsec} \\
f &= 0.15 \\
\beta &= 1.5 \\
k_z &= 0.7 \\
z_m &= 0.47 \\
z'_m &= 0.1
\end{aligned} \qquad (3.22)$$

In this model, the median redshift for an $r = 22$ galaxy is $z = 0.47$, in rough agreement with observations. At this redshift $\theta_{\max}$ corresponds to $118 h^{-1}$ kpc. Using this model, then, we have evaluated Eq. (3.21) and show the results as a contour plot of $\langle p \rangle$ as a function of $r_d, r_s$ in Fig. 4.

The image polarisation measured from an observational sample is, of course, an average over a distribution of source magnitudes, $r_s$, and lens magnitudes, $r_d$. Assuming the number counts of galaxies as a function of $r$-magnitude follow a relation of the form $\frac{dN}{dr} = A \, 10^{0.33 r}$, we find the mean image polarisation based upon our analytic model to be $\langle p \rangle = 0.011$ for lenses and sources corresponding to our principal observational sample ($20 \leq r_d \leq 23$ and $23 < r_s \leq 24$).

### 3.4 Calculation of Polarisation

Using linear theory, we can predict the distribution in image polarisation. As in §2.1 we write the complex orientation of the image of a source galaxy as $\chi = \epsilon e^{2i\phi}$. If the polarisation is $p$, then to linear order the intrinsic source orientation is $\chi_0 = \chi - p$ and the effect of the lens galaxy is a simple translation on the $\chi$ plane. Let this translation be in the $x$ direction so that the normalised, observed distribution of orientations is

$$f_\chi(\chi_x, \chi_y) = f_{\chi 0}(\chi_x - p, \chi_y) \tag{3.23}$$

where $f_{\chi 0}$ is the intrinsic distribution in $\chi_0$. An unbiased estimator of the polarisation is then given using

$$\begin{aligned}
\langle \chi \rangle &= \int d\chi_x d\chi_y f_\chi(\chi_x, \chi_y) \chi_x \\
&= \int d\chi_{0x} d\chi_{0y} f_{\chi 0}(\chi_{0x}, \chi_{0y}) \chi_{0x} + p \int d\chi_x d\chi_y f_\chi(\chi_x, \chi_y) \\
&= p \int d\chi_x d\chi_y f_\chi(\chi_x, \chi_y)
\end{aligned} \tag{3.24}$$

where we have assumed that $f_{\chi 0}$ is isotropic. It is cleanest to estimate $p$ directly using

$$p = \frac{\int d\chi_x d\chi_y f_\chi(\chi_x, \chi_y) \chi_x}{\int d\chi_x d\chi_y f_\chi(\chi_x, \chi_y)} \tag{3.25}$$

However, what we actually measure is the distribution in $P_\phi(\phi)$ integrated over ellipticity. We therefore write

$$\begin{aligned}
P_\phi(\phi) &= \int d\epsilon \; \epsilon f_{\chi 0} + p \cos 2\phi \int d\epsilon \; \epsilon \frac{df_{\chi 0}}{d\epsilon} \\
&= \left(\frac{2}{\pi}\right) \left[1 - \langle p \rangle \cos 2\phi \left\langle \epsilon^{-1} \right\rangle \right]
\end{aligned} \tag{3.26}$$

where the best fit ellipticity distribution, Eq. (2.1), yields $\langle \epsilon^{-1} \rangle = 8.0$ and $\langle p \rangle$, the average polarisation in the sample, should vary with the lens and source galaxy selection criteria.

### 3.5 Monte Carlo Simulations

To determine the best-fit halo parameters for our lensing model we have constructed Monte Carlo simulations of gravitational lensing of background galaxies by foreground galaxies. For every observed object with $20 \leq r \leq 24$, we assign a galaxy to a random location on the Monte Carlo CCD frame (the angular clustering of the observed objects is weak, eg. Brainerd, Smail & Mould (1995), and has a negligible effect on the predicted signal). The galaxies are then assigned $r$-magnitudes and ellipticities drawn at random from the corresponding observed distributions and position angles uniformly distributed on $[-\pi/2, \pi/2]$. Based on its assigned $r$-magnitude, each galaxy is then given a redshift chosen from the parameterised redshift distribution given by Eq. (3.17), where the parameters assumed for the redshift distribution are those summarised by Eq. (3.22).

The mass distribution of the lens galaxies is modeled according to Eq. (3.1) and we assume all lenses follow same scaling relations with $V^*$ and $s^*$ (i.e. Eq. (3.10) and Eq. (3.11)), which are the parameters we wish to investigate. For each source galaxy we compute the net image polarisation induced by all foreground lens galaxies. Since we are dealing with the weak lensing limit, it is sufficient to compute the individual image polarisations due to distinguishable lenses and superpose them to obtain the net polarisation. For the adopted galaxy redshift distributions we find that roughly a third of the sources are lensed by only a single foreground galaxy, another third are affected by two lenses and the remaining third encounter 3 or more significant deflectors.

Having computed the net polarisation of each of the faint galaxy images, we then compute $P_\phi(\phi)$ for the Monte Carlo images in exactly the same manner as the actual data shown in Fig. 2a. By performing many ($\sim 1000$) Monte Carlo realisations for a given pair of scaling parameters $(V^*, s^*)$, we determine a good estimate of the faint galaxy polarisation for a particular lensing model. By comparing the predicted and observed $\langle p \rangle$ and using a $\chi^2$ minimisation, we can then obtain best-fit parameters for the dark halos of the lenses. Results of this procedure are summarised in §4.

### 3.6 Sensitivity to model parameters

As mentioned above, the predicted image polarisation is simply proportional to the square of the assumed fiducial circular velocity $V^*$. However, the dependence upon the outer radial scale, $s^*$, is more subtle. If we increase $s^*$ from its fiducial value to $\simeq 300 h^{-1}$ kpc, keeping all other parameters in the standard model the same, the polarisation remains constant at 0.011. Only if we reduce $s^*$ substantially is there a substantial reduction in the measured polarisation. For example, changing $s^*$ to $30 h^{-1}$ kpc reduces $\langle p \rangle$ from 0.011 to 0.009. The reason for this behaviour is that, within our standard model, most of the signal is contributed by lenses that are sufficiently close to the source on the sky that the line of sight passes through the isothermal part of the dark halo and, as $p^* \propto s^{-1}$ for a given velocity dispersion and $\bar{G} \propto s$, for large halos the average polarisation is approximately independent of $s$. The overall mass to light ratio will, however, increase in proportion to $s$.

Next we vary the spectral index, $\alpha$, independently of the other parameters. Changing $\alpha$ from 3 to 2 reduces $\langle p \rangle$ from 0.011 to 0.009. This comparatively weak variation indicates that our simplistic model for the K-correction is not a serious concern.

We now determine the sensitivity of the model predictions to the lens and source redshift distributions. Let us keep the median redshift (and its $z$ variation) constant but change the shape of the distribution. Reducing the high $z$ tail by increasing $\beta$ from 1.5 to 2 (a Gaussian) results in a reduction of the mean polarisation from 0.011 to 0.008.

Alternatively, we can keep the shape of the redshift distribution unchanged and increase the median redshift. In this case, we find that shifting $z_m$ from 0.47 to 0.7 results in a mean polarisation of 0.015. It turns out that our choice of variation of median redshift for the source galaxy distribution with magnitude more or less maximises the predicted polarisation and there is only a small sensitivity to $z'_m$.

Similarly, there is a little sensitivity to the world model. For example, changing from an Einstein-De Sitter universe to an empty universe ($\Omega_0 = 0$, $D_{ds} = c(2H_0 a_d)^{-1}(2 + a_s + a_d)(a_d - a_s)$) results in a mean polarisation of 0.012.

Our measurement of the mean image polarisation used a fixed aperture around each lens galaxy of 34 arcsec. It turns out that this, too, roughly maximises the signal to noise in our simulations. To understand this, first observe that the induced polarisation is linear in the mass distribution and so it ought not to matter if several lens galaxies are contributing as long as they can be treated as randomly oriented. If we increase $\theta_{\max}$, then the number of lens-source pairs will increase as $\theta_{\max}^2$, and the random error will diminish as $\theta_{\max}^{-1}$. The polarisation signal from these extra background galaxies will diminish as $\theta_{\max}^{-1}$ as long as we are still looking through the dark halos, with $\theta_{\max} \lesssim \theta_c$ and thus the signal to noise ought to improve logarithmically with increasing $\theta_{\max}$. However, when the impact parameters exceed the lens outer radii, $s$, the signal to noise will deteriorate. Consequently, there is little gain in signal to noise possible from increasing the aperture size beyond $s$. For example, increasing $\theta_{\max}$ from 34 to 60 arcsec, leads to a reduction in the mean polarisation signal $\langle p \rangle$ from 0.011 to 0.006, a reduction by a factor 1.8, even though the number of galaxies pairs increases by a factor 3.1, which should reduce the noise by a factor 0.57, leaving the signal to noise ratio constant. Conversely, reducing $\theta_{\max}$ to 15 arcsec, increases $\langle p \rangle$ by a factor of 2.2, but reduces the number of pairs by 0.18 so that the signal to noise diminishes by 0.8.

## 4. RESULTS

Using the functional form for $P_\phi(\phi)$ given by Eq. (3.26) and fitting to the observations we obtain $\langle p \rangle = 0.011 \pm 0.006$ (95% confidence bounds) for our fiducial galaxy sample ($20.0 \leq r_d \leq 23.0$, $23.0 < r_s \leq 24.0$) and $5 \leq \theta \leq 34$ arcsec. As discussed in §2, we cannot create such a signal from systematic effects in either our dataset or our analysis. We must therefore accept that the signal is real and search for an astrophysical origin. In this regard, two processes present themselves as obvious contenders to produce a preferential alignment of faint galaxy images around brighter systems: tidally distorted companions or gravitational lensing.

We now show that companion galaxies are unlikely to be the source of the observed signal. Limits can be set on the magnitude of the contamination due to tidally-induced tangential elongations of genuine satellite galaxies of the lens galaxies (cf. Phillips 1985, Tyson 1985) from the clustering strength of the galaxies. Fortunately, the clustering statistics of the galaxies in this field have already been carefully determined (Brainerd, Smail & Mould 1995). We estimate the angular cross correlation function for galaxies in our primary data set with $20 \leq r_d \leq 23$, $23 < r_s \leq 24$ to be

$$w(\theta) = 0.6 \left(\frac{\theta}{1''}\right)^{-0.9}, \qquad (4.1)$$

where we have computed the integral constraint directly for our frame and allowed a 15% stellar contamination of the object catalogue (see Brainerd, Smail & Mould 1995). Let us suppose that that the mean polarisation associated with intrinsic tidal effects is $p_t$. Our cursory examination of images of nearby galaxies reveals that $|p_t| < 0.1$ for satellites within a magnitude fainter of their neighbour and separations $\sim 100$ kpc. The contribution of the tidal elongation to the observed polarisation signal should therefore be of order $1.6w(\theta_{\max})p_t$, which is smaller in modulus than 0.004 for $\theta_{\max} = 34$ arcsec and of order $2.3\sigma$ below our measured image polarisation. We therefore conclude that a contribution to the observed polarisation due to a population of tidally distorted dwarfs can be ruled out with moderate confidence. A contrary calculation that illustrates the effect of gravitational lensing on the measured autocorrelation function of galaxies at faint magnitudes is given in Appendix B.

We conclude that the most probable cause of the signal reported in §2 is the distortion of the distant galaxy images due to gravitational lensing by foreground galaxies. Our fiducial model, Eq. (3.22), yields a prediction of $\langle p \rangle = 0.011$ for the analytic model and $\langle p \rangle = 0.009 \pm 0.003$ for the Monte Carlo simulations. The theoretical predictions of $\langle p \rangle$ are, thus, in good agreement with each other and with the observational measurement. We conclude that we are able to recreate easily the observed lensing signal with a simple, physically motivated model, strengthening the case for galaxy–galaxy lensing as the cause of the observed image polarisation.

To determine the level at which we can constrain the parameters of our lensing model, we have investigated the radial dependence of the polarisation signal. The image polarisation will diminish for lines of sight passing outside the individual dark matter halos of the lenses and, so, the scale at which the observed $\langle p \rangle$ approaches zero is an indication of the angular extent of the halos. In Fig. 5a we show $\langle p \rangle$ for our fiducial subsample of galaxies calculated in annuli of inner radius 5 arcsec and outer radius $\theta_{\max}$, where $15 \leq \theta_{\max} \leq 145$ arcsec. These measurements of $\langle p \rangle$ are, of course, not independent but they do serve to illustrate that $\langle p \rangle$ is significantly non-zero up to scales of $\theta_{\max} \sim 90$ arcsec, suggesting the dark halos of the lenses extend to large radii ($\gtrsim 100h^{-1}$ kpc).

A more ambitious approach is possible if we attempt to measure the sizes of the lens galaxies by determining the variation of polarisation with the differential source–lens separation, $\theta$. Again using our fiducial subsample of galaxies, we have computed $\langle p \rangle$ as a function of $\theta$ and the results are exhibited in Fig. 5b. In addition, we have computed the mean polarisation of the faint galaxy images in the Monte Carlo simulations as a function of $\theta$ for our fiducial $L^*$ galaxy model assuming different values of the outer radial scale, $s^*$. Results for $V^* = 220$ km s$^{-1}$ and $s^* = 20h^{-1}$ kpc, $100h^{-1}$ kpc, $250h^{-1}$ kpc are shown in Fig. 5b.

Keeping $V^*$ constant at 220 km s$^{-1}$, $s^*$ was increased incrementally from $10h^{-1}$ kpc to $250h^{-1}$ kpc in the Monte Carlo simulations and the variation of $\langle p \rangle$ with $\theta$ was determined for each value of $s^*$. A $\chi^2$ statistic comparing the model predictions to the observed $\langle p \rangle$ in Fig. 5b was then computed. As $s^*$ was increased from $10h^{-1}$ kpc, $\chi^2$ decreased monotonically from $\sim 7.5$ and reached a constant value of $\sim 3.0$ for $s^* \gtrsim 100h^{-1}$ kpc. That is, for values of $s^* \gtrsim 100h^{-1}$ kpc, the expected variation of $\langle p \rangle$ is essentially constant, as discussed in §3.6. We are, therefore, unable to determine a unique best-fit value of $s^*$, but the $\chi^2$ minimisation suggests that the halos of the lenses would be at least $100 h^{-1}$ kpc in radial extent.

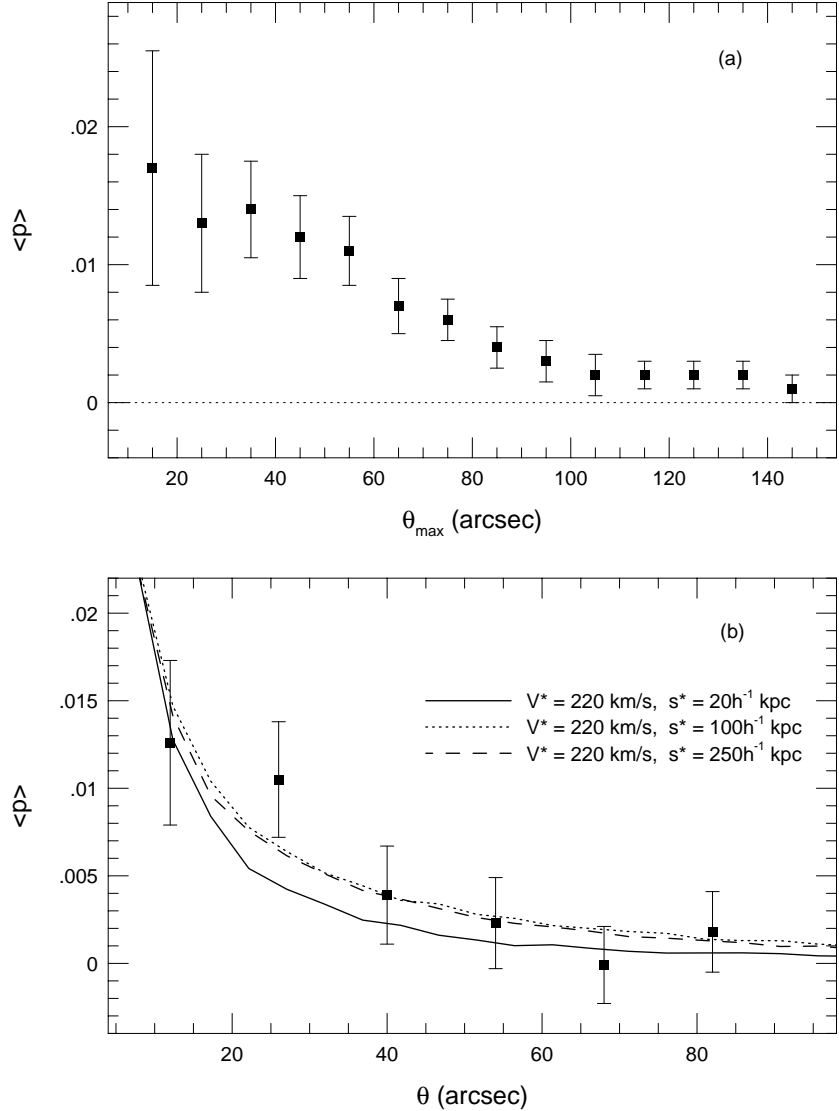

Fig. 5: Angular variation of image polarisation for foreground galaxies with $20 \leq r_d \leq 23$ and background galaxies with $23 < r_s \leq 24$. (a) Variation of $\langle p \rangle$ with increasing annulus outer radius, $\theta_{\max}$. (b) Variation of $\langle p \rangle$ with differential lens–source separation, $\theta$. Theoretical estimates of $\langle p \rangle$ for fiducial $L^*$ galaxy gravitational lenses (see §3) with different scaling radii, $s^*$, are also shown.

It is clear from Fig. 5b that our fiducial model with $s^* \gtrsim 100 h^{-1}$ kpc yields a good fit to the observed variation of $\langle p \rangle$ with source–lens separation. However, it is also interesting to investigate the amount by which we can change $V^*$ and still obtain a reasonable $\langle p \rangle$ compared to the observations. Keeping $s^*$ constant at $100 h^{-1}$ kpc, $V^*$ was varied in the Monte Carlo simulations and, as above, a $\chi^2$ statistic was used to compare the observed variation of $\langle p \rangle$ with $\theta$ to the model predictions. From this we find the upper and lower 90% confidence bounds on $V^*$ to be $\sim 304$ km s$^{-1}$ and $\sim 116$ km s$^{-1}$.

Finally, it is clear from Eq. (3.17) that the redshift distributions of our fiducial subsamples of lenses ($20 \leq r_d \leq 23$) and sources ($23 < r_s \leq 24$) will have some overlap. We expect the measured mean polarisation to be greatest for the most distant sources and for the case that all the sources are at higher redshifts than the candidate lenses. We are therefore driven to look at using additional information to better separate the foreground lenses from the background sources. One obvious possibility is to use the colours of the faint sources from our matched $g$ and $r$-band data. We thus split the source sample with $23 < r_s \leq 24$ into a "red" half $[(g-r) > 0.53]$ and a "blue" half $[(g-r) < 0.53]$ on the basis of their $(g-r)$ colours, and compute $\langle p \rangle$ for the 2 subsamples. Again using $5 \leq \theta \leq 34$ arcsec, we find $\langle p \rangle_{\text{blue}} = 0.016 \pm 0.008$ and $\langle p \rangle_{\text{red}} = 0.008 \pm 0.008$ (95% confidence limits). In addition, we compute $\langle p \rangle$ for the red and blue subsamples as in Fig. 5a and in Fig. 6 we show the measured variation of $\langle p \rangle$ with $\theta_{\max}$. From this figure $\langle p \rangle_{\text{red}}$ appears consistent with zero over all scales, while $\langle p \rangle_{\text{blue}}$ is significantly non-zero for $\theta_{\max} \lesssim 60$ arcsec and there is evidence of a monotonic decrease to zero with increasing $\theta_{\max}$. While not highly significant, these data suggest a higher mean polarisation of the blue images than the red images, a result which makes sense if a larger proportion of distant sources are blue, star-forming systems.

## 5. DISCUSSION

In this paper we have attempted to measure the induced polarisation of images of distant galaxies due to weak gravitational lensing by more nearby galaxies. We have a significant detection of this polarisation of $\langle p \rangle = 0.011 \pm 0.006$ (95% confidence bounds). We cannot explain this signal through systematic effects within our dataset and thus we believe it is real. In addition, we have presented a fiducial model which is capable of reproducing the observed gravitational lensing signal though both analytic and Monte Carlo simulations. The simulation results are quite robust to most variations apart from the scaling circular velocity, $V^*$, and the details of the redshift distribution. With somewhat lower significance we can claim to have measured the decrease in the signal with increasing galaxy separation, consistent with a typical galaxy halo size $\gtrsim 100 h^{-1}$ kpc.

The best-fitting model parameters from the Monte Carlo simulations can be used to estimate the masses of the lens galaxies contained within a radius, $r$. For $V^* = 220 \text{km s}^{-1}$ and $s^* = 100 h^{-1}$kpc, we find $M(100 h^{-1}\text{kpc}) = 8.8 \times 10^{11} h^{-1} M_\odot$. For the 90% confidence limits on $V^*$ derived above this becomes $M(100 h^{-1}\text{kpc}) = (8.8^{+8.2}_{-6.3}) \times 10^{11} h^{-1} M_\odot$. Letting $s^* \to \infty$, we obtain a maximum contained mass of $M_{\max}(100 h^{-1}\text{kpc}) = (1.1^{+1.0}_{-0.8}) \times 10^{12} h^{-1} M_\odot$ for our allowed range of $V^*$. From our model calculations, then, we estimate an allowed range for the masses of the lens halos to be: $M \sim 1.0^{+1.1}_{-0.7} \times 10^{12} h^{-1} M_\odot$.

The typical luminosities of the lens galaxies in our sample, given our fiducial median redshift, are $L_V \sim 5 \times 10^9 h^{-2} L_\odot$. We thus estimate rest-frame mass to light ratios inside a radius of $100 h^{-1}$ kpc of our lensing galaxies to to be $M/L_V = 200^{+220}_{-140} h (M/L_V)_\odot$ for $\Omega_0 = 1$. A value of $M/L_V = 1400 h (M/L_V)_\odot$ is required for closure density and we, therefore, estimate the fraction of the closure density contained in the central regions of galaxies to be $\Omega_g = 0.14^{+0.16}_{-0.10}$. Our observations thus provide the first evidence from weak gravitational lensing of large-scale dark halos associated with individual galaxies.

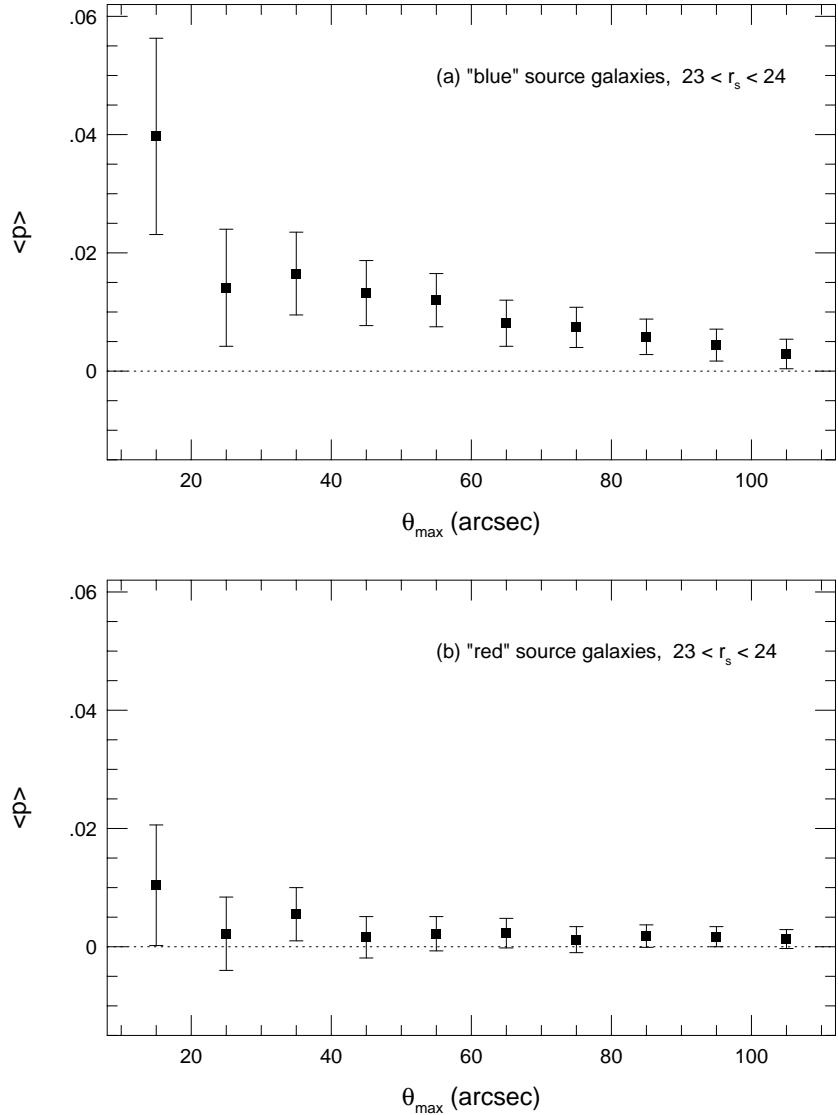

Fig. 6: Angular variation of image polarisation for the "blue", $(g - r) < 0.53$, and "red", $(g - r) > 0.53$, source subsamples as a function of outer annular radius, $\theta_{\max}$, where $23 < r_s \leq 24$.

Few measurements exist for galaxy masses on the scales probed here. Perhaps the best existing estimates of the extent and masses of galaxy halos come from statistical studies of satellite galaxies. Zaritsky & White (1994) have analysed a sample of companion galaxies to isolated local spirals and find that the typical masses out to $150h^{-1}$ kpc are in the range $1$–$2 \times 10^{12} h^{-1} M_\odot$, consistent with our findings, $M(150h^{-1}\mathrm{kpc}) = 1.4^{+1.8}_{-1.1} \times 10^{12} h^{-1} M_\odot$. We stress that the assumptions underlying the two techniques are very different which makes the close agreement from the two methods very encouraging.

We conclude that weak gravitational lensing is a viable and potentially powerful probe of the outer parts of normal galaxies – regions that are inaccessible to strong lensing studies (Breimer & Sanders 1993). That our polarisation measurement has been achieved using only a single CCD frame under conditions of modest seeing augurs well for future investigations. When planning such observation a compromise has to be reached between area coverage and depth when accruing the source and lens samples. Galaxy counts increase at a rate $\sim 2$ per magnitude, hence the number of galaxy-lens pairs should quadruple for each magnitude and, theoretically, the random error in the polarisation measurement should halve, making it advantageous to go fainter. In practice, the limiting magnitude is set by our ability to assign accurate shapes to individual source galaxies and we have seen that the empirical limit for our data is $r \sim 24$ (Fig. 2). Note that, in the absence of scale evolution, beyond a redshift $z \sim 0.5$ the image sizes of galaxies are relatively fixed and to increase the depth of the sample by $\sim$ one magnitude takes $\sim 5$ times as much integration. Consequently, even when the galaxy images can be measured accurately, galaxy-lens pairs are accumulated at about the same rate by taking additional CCD frames as by increasing the depth of exposure of an individual image.

It is clear from our observations and simulations that atmospheric seeing seriously degrades the polarisation at faint magnitudes where the signal to noise would otherwise increase. This should not be a problem for deep HST images and it is important to repeat this measurement using the deepest WFPC2 fields as they become publicly available. Using the archival data we anticipate being able to measure image ellipticities accurately to $r = 26$. In this case, the lensing signal should go up by a factor of 3, while the noise should go down by a factor of 4 in a given area. Currently, on the order of 10 fields should be appropriate for our analysis, so that the signal to noise may improve by a factor of 30 (although this will depend upon the galaxy redshift distribution). It should, therefore, be possible to improve the accuracy of the measurement substantially and constrain both the redshift distribution of faint galaxies and the sizes of their halos.

The eventual limitations to this technique are probably connected with our incomplete knowledge of the galaxy redshift distribution and the unknown contribution to the statistical signal of a minority of unusual lenses and tidal interactions. Despite these concerns, the prospect of studying galaxies on a scale where their dynamical times are comparable to their ages is exciting. In this regime we can observe the infall of the outer parts of the galaxies and thus, in some sense, see the galaxies assembling. This encourages us to devote more observational effort to measuring galaxy-induced polarisation.

## ACKNOWLEDGMENTS

We are indebted to Jeremy Mould and Todd Small for acquiring the observations used for this analysis and to them, David Hogg, and Tony Tyson for helpful suggestions. Support under NSF contract AST 92-23370, the NASA HPCC program at Los Alamos National Laboratory (TGB) and a NATO Advanced Fellowship (IRS) is gratefully acknowledged.

# REFERENCES

Aaronson, M. & Mould, J. R. 1983, ApJ, 265, 1

Binney, J. J. & Tremaine, S. D. 1987, Galactic Dynamics Princeton: Princeton University Press

Brainerd, T. G., Smail, I. & Mould, J. R. 1995, MNRAS, in press

Blandford, R. D. & Narayan, R. 1992, ARAA, 30, 311

Breimer, T. G. & Sanders, R. H. 1993, AA, 274, 96

de Zeeuw, T. & Franx, M. 1991, ARAA, 29, 239

Dressler, A. et al. 1995, in prep

Efstathiou, G., Bernstein, G., Katz, N., Tyson, J. A., & Guhathakurta, P. 1991, ApJ, 380, L47

Fabbiano, G. 1989, ARAA, 27, 87

Fich, M. & Tremaine, S. D. 1991, ARAA, 29, 409

Fort, B. & Mellier, Y. 1994, Astron. Astrophys. Rev., 5, 239

Kovner, I, & Milgrom, M. 1987, ApJ, 321, L113

Kristian, J. 1967, ApJ, 147, 864

Kristian, J. & Sachs, R. K. 1966, ApJ, 143, 379

Lilly, S. 1993, ApJ, 411, 501

Loveday, J., Peterson, B. A., Efstathiou, G., Maddox, S. J. & Sutherland, W. J. 1992, ApJ, 390, 338

Mould, J., Blandford, R., Villumsen, J., Brainerd, T., Smail, I., Small, T., & Kells, W. 1994, MNRAS, 271, 31

Narayan, R. 1989, ApJ, 339, L53

Phillips, S. 1985, Nature, 314, 721

Schneider, P., Ehlers, J. & Falco, E. 1992, Gravitational Lenses Berlin: Springer-Verlag

Tresse, L., Hammer, F., LeFevre, O. & Proust, D. 1993, A&A, 277, 53

Tyson, J. A., Valdes, F., Jarvis, J. F. & Mills, A. P. 1984, ApJ, 281, L59

Tyson, J. A. 1985, Nature, 316, 799

Valdes, F. 1982, FOCAS Manual, NOAO,

Valdes, F., Tyson, J. A. & Jarvis, J. F. 1983, ApJ, 271, 431

Webster, R. L. 1983, PhD Thesis, University of Cambridge

Zaritsky, D. & White, S. D. M. 1994, ApJ, 435, 599
2323# REFERENCES

Aaronson, M. & Mould, J. R. 1983, ApJ, 265, 1

Binney, J. J. & Tremaine, S. D. 1987, Galactic Dynamics Princeton: Princeton University Press

Brainerd, T. G., Smail, I. & Mould, J. R. 1995, MNRAS, in press

Blandford, R. D. & Narayan, R. 1992, ARAA, 30, 311

Breimer, T. G. & Sanders, R. H. 1993, AA, 274, 96

de Zeeuw, T. & Franx, M. 1991, ARAA, 29, 239

Dressler, A. et al. 1995, in prep

Efstathiou, G., Bernstein, G., Katz, N., Tyson, J. A., & Guhathakurta, P. 1991, ApJ, 380, L47

Fabbiano, G. 1989, ARAA, 27, 87

Fich, M. & Tremaine, S. D. 1991, ARAA, 29, 409

Fort, B. & Mellier, Y. 1994, Astron. Astrophys. Rev., 5, 239

Kovner, I, & Milgrom, M. 1987, ApJ, 321, L113

Kristian, J. 1967, ApJ, 147, 864

Kristian, J. & Sachs, R. K. 1966, ApJ, 143, 379

Lilly, S. 1993, ApJ, 411, 501

Loveday, J., Peterson, B. A., Efstathiou, G., Maddox, S. J. & Sutherland, W. J. 1992, ApJ, 390, 338

Mould, J., Blandford, R., Villumsen, J., Brainerd, T., Smail, I., Small, T., & Kells, W. 1994, MNRAS, 271, 31

Narayan, R. 1989, ApJ, 339, L53

Phillips, S. 1985, Nature, 314, 721

Schneider, P., Ehlers, J. & Falco, E. 1992, Gravitational Lenses Berlin: Springer-Verlag

Tresse, L., Hammer, F., LeFevre, O. & Proust, D. 1993, A&A, 277, 53

Tyson, J. A., Valdes, F., Jarvis, J. F. & Mills, A. P. 1984, ApJ, 281, L59

Tyson, J. A. 1985, Nature, 316, 799

Valdes, F. 1982, FOCAS Manual, NOAO,

Valdes, F., Tyson, J. A. & Jarvis, J. F. 1983, ApJ, 271, 431

Webster, R. L. 1983, PhD Thesis, University of Cambridge

Zaritsky, D. & White, S. D. M. 1994, ApJ, 435, 599
23

**APPENDIX A: Particle distribution function associated with model potential**

In Sec. 3.1, we introduced a simple galaxy model with a density distribution given by Eq. (3.1). Associated with this density distribution is the 3D potential

$$\phi(r) = -\int_r^\infty dr' \frac{GM(r')}{r'^2} = -V_c^2 \left[ \frac{\tan^{-1} x}{x} + \frac{1}{2} \ln(1 + x^{-2}) \right], \quad (A1)$$

where $x = r/s$. In order to demonstrate that this potential is physically realisable, we compute the associated isotropic distribution function for dark matter particles assuming that they are all of the same mass $m$ and ignoring the luminous stars, which ought only to contribute at small radius. We denote this distribution function by $f(E)$, where $E = v^2/2 + \phi$ is the specific energy. If we now regard $\rho$ as a function of $\phi$, then we can write

$$\rho(\phi) = 4\,2^{1/2}\pi m \int_\phi^0 dE (E - \phi)^{1/2} f(E), \quad (A2)$$

eg. Binney & Tremaine (1987) This integral equation is easily cast in Abel form and can be solved to give

$$f(E) = \frac{1}{2^{1/2}\pi^2 m} \int_E^0 d\phi (\phi - E)^{1/2} \frac{d^3\rho}{d\phi^3} \quad (A3)$$

where we have imposed the boundary condition that $\rho$ and its derivatives $\to 0$ as $\phi \to 0$. This distribution function can be evaluated numerically and is found to decline monotonically with $E$. In addition, $d^3\rho/d\phi^3 < 0$. These properties suffice to ensure stability to small perturbations (eg. Binney & Tremaine 1987).

# APPENDIX B: Influence of weak lensing on galaxy autocorrelation function

Curiously, weak gravitational lensing produces an observable effect on the autocorrelation function of distant galaxies. It might be thought that the magnification of more distant galaxies by intervening lenses would result in a positive contribution to the measured autocorrelation function. However, the deflection of the galaxy images also expands the separation between galaxies and this turns out to be the larger effect. If the slope of the galaxy counts can be expressed as

$$\frac{d \log N}{dr} = q \qquad (B1)$$

where $q \sim 0.3$, then the surface density of background galaxies at a distance $\theta$ from a given lens galaxy will be given be enhanced by a factor $1 + w'(\theta)$ where

$$w'(\theta) \simeq (2.5q - 1)\mu(\theta) \qquad (B2)$$

and $\mu(\theta)$ is the magnification (eg. Narayan 1989). The contribution of gravitational lensing to the correlation function turns out to be negative, reducing the very small scale clustering since the distance between galaxies has been increased. We can relate $w'(\theta)$ to the polarisation, $p(\theta)$, using Eq. (3.2),(3.7), and we find

$$\frac{w'(\theta)}{p(\theta)} = (2.5q - 1)\left(\frac{1 - (1 + X^{-2})^{-1/2}}{XG(X)}\right). \qquad (B3)$$

This increases from $\sim -1/4$ for $X \ll 1$ to $\sim -1/16X$ for $X \gg 1$. This ratio is always small and so cannot affect the measured image polarisation. However, it can have an effect upon the autocorrelation function of galaxies fainter than those scrutinised here by canceling a significant fraction of the true correlation on scales $\lesssim 10$ arcsec. On scales of $\sim 15$–$20$ arcsec, however, the effect is very small and it is negligible on scales $\gtrsim 50$ arcsec. Therefore, weak gravitational lensing cannot account for the observed low clustering amplitude of very faint galaxies on scales $\gtrsim 15$ arcsec (eg. Brainerd, Smail & Mould, 1995; Efstathiou et al. 1991).

TABLE 1

| faint image magnitudes | number of faint images | number of faint–bright pairs | $\chi^2$ rejection confidence level | KS rejection confidence level |
| --- | --- | --- | --- | --- |
| $23.0 < r \leq 24.0$ | 506 | 3202 | 98.6% | 99.9% |
| $23.0 < r \leq 25.0$ | 1755 | 10870 | 97.3% | 99.2% |
| $23.0 < r \leq 26.0$ | 4303 | 26412 | 19.4% | 60.0% |

**FIGURE CAPTIONS**

Fig. 1. Orientation of faint galaxies relative to bright galaxies.

Fig. 2. Probability distribution $P_\phi(\phi)$ of orientation of faint galaxies relative to the directions of bright galaxies with projected separations $5 \leq \theta \leq 34$ arcsec. The bright galaxies have $20 \leq r \leq 23$ and the faint galaxies have magnitudes in the ranges indicated in the text and figure panels. For the best case, (a), of a non-uniform $P_\phi(\phi)$, the best fitting theoretical $\cos 2\phi$ variation is also shown.

Fig. 3. Scaled polarisation variation with angle $\theta$ for a given galaxy. $G(X)$ is defined in Eq. (3.9) and $X = R/s$.

Fig. 4. Theoretical variation of polarisation as a function of source magnitude $r_s$ and lens galaxy magnitude $r_d$ according to Eq. (3.21).

Fig. 5. Angular variation of image polarisation for foreground galaxies with $20 \leq r_d \leq 23$ and background galaxies with $23 < r_s \leq 24$. (a) Variation of $\langle p \rangle$ with increasing annulus outer radius, $\theta_{\max}$. (b) Variation of $\langle p \rangle$ with differential lens–source separation, $\theta$. Theoretical estimates of $\langle p \rangle$ for fiducial $L^*$ galaxy gravitational lenses (see §3) with different scaling radii, $s^*$, are also shown.

Fig. 6. Angular variation of image polarisation for the "blue", $(g - r) < 0.53$, and "red", $(g - r) > 0.53$, source subsamples as a function of outer annular radius, $\theta_{\max}$, where $23 < r_s \leq 24$.